\documentstyle[ApJ,psfig]{article}
\tighten

\lefthead{Woods et al.}
\righthead{Discovery of SGR~1627--41}

\slugcomment{To be submitted to ApJ Letters}

\begin{document}

\title{Discovery of a New Soft Gamma Repeater, SGR~1627--41}

\author{
Peter~M.~Woods\altaffilmark{1},
Chryssa~Kouveliotou\altaffilmark{2,3},
Jan~van~Paradijs\altaffilmark{1,4},
Kevin~Hurley\altaffilmark{5},
R.~Marc~Kippen\altaffilmark{3,6},
Mark~H.~Finger\altaffilmark{2,3},
Michael~S.~Briggs\altaffilmark{1,3},
Stefan~Dieters\altaffilmark{1,3} and
Gerald~J.~Fishman\altaffilmark{3}
}

\altaffiltext{1}{Deptartment of Physics, University of Alabama in Huntsville, 
Huntsville, AL 35899; peter.woods@msfc.nasa.gov}
\altaffiltext{2}{Universities Space Research Association}
\altaffiltext{3}{NASA Marshall Space Flight Center, ES--84, Huntsville, AL
35812}
\altaffiltext{4}{Astronomical Institute ``Anton Pannekoek'', University of 
Amsterdam, 403 Kruislaan, 1098 SJ Amsterdam, NL}
\altaffiltext{5}{University of California, Berkeley, Space Sciences Laboratory,
Berkeley, CA 94720-7450}
\altaffiltext{6}{CSPAR (Center for Space Plasma, Aeronomic and Astrophysics
Research), University of Alabama in Huntsville, Huntsville, AL 35899}

\begin{abstract}

We report the discovery of a new soft gamma repeater (SGR), SGR~1627--41, and
present   BATSE observations of the burst emission and BeppoSAX NFI
observations of the probable persistent X-ray counterpart to this SGR.  All but
one burst spectrum are well fit by an optically thin thermal bremsstrahlung
(OTTB) model with $kT$ values between 25 and 35 keV.  The spectrum of the X-ray
counterpart, SAX~J1635.8--4736, is similar to that of other persistent SGR
X-ray counterparts.  We find weak evidence for a periodic signal at 6.41 s in
the light curve for this source.  Like other SGRs, this source appears to be
associated with a young supernova remnant G337.0--0.1.  Based upon the peak
luminosities of bursts observed from this SGR, we find a lower limit on the
dipole magnetic field of the neutron star B$_{\rm dipole}~\gtrsim
5~\times~10^{14}$ Gauss. 

\end{abstract}

\section{INTRODUCTION}

Soft gamma repeaters (SGRs) are a rare type of stellar object characterized by
their transient emission of bursts of hard X-rays and soft $\gamma$-rays. 
Bursts have been detected from three such sources from 1979 (Mazets et al.
1981) until early 1998; two are in the galactic plane (SGR~1806--20,
SGR~1900+14) and one is in the Large Magellanic Cloud (SGR~0526--66).  One of
the first SGR bursts  detected, the famous 1979~March~5 burst from SGR~0526--66
(Mazets et al. 1979), provided a wealth of information about these sources
(Thompson \& Duncan 1995).  This flare started with a short initial spike
followed by a 3 minute train of coherent 8 s pulsations.  A precise location of
this burst was consistent with a young ($\sim$ 10$^4$ year) supernova remnant
(SNR) N49 (Cline et al. 1982).  The train of pulsations and the positional
coincidence with the SNR indicated that the burst source is a young, magnetized
neutron star.

Pointed X-ray observations of SGR burst location regions have shown that each
SGR has associated with it a persistent X-ray source (Murakami et al. 1994,
Hurley et al. 1999a, Rothschild et al. 1994) within or near a young SNR
(Kulkarni \& Frail 1993, Hurley et al. 1999a, Cline et al. 1982).  Furthermore,
the persistent sources associated with the two galactic SGRs are X-ray pulsars
(Kouveliotou et al. 1998a, Hurley et al. 1999a) which show secular spin down at
a rate $\sim$ 10$^{-10}$ s s$^{-1}$ (Kouveliotou et al. 1998a, 1999).  As
argued by Kouveliotou et al. (1998a), this spin down is likely caused by
magnetic dipole radiation which implies a neutron star dipole magnetic field of
$\sim$ 10$^{14 - 15}$ Gauss.  These results have provided strong observational
evidence  in support of the idea that SGRs are strongly magnetized neutron
stars or `magnetars' (Duncan \& Thompson 1995).    

The majority of SGR bursts have durations less than 200 msec and are well
characterized by optically thin thermal bremsstrahlung (OTTB) spectra with
temperatures $kT$ $\sim$ 30 -- 40 keV (Kouveliotou 1995).  With the exception
of the much more luminous March 5$^{th}$ event and a similar bright flare
detected recently from SGR~1900+14 (Hurley et al. 1999c), SGR bursts reach peak
luminosities up to $\sim$ 10$^{42}$ ergs sec$^{-1}$, far exceeding the
Eddington luminosity for a 1.4 $M_{\odot}$ neutron star.  A statistical study
of bursts from SGR~1806--20 has shown that no correlation exists between the
energy released in a burst and the time until the next burst (Laros et al.
1987).  Also, it was found that both burst peak fluxes and time intervals
between bursts resemble truncated log-normal and log-normal distributions,
respectively (Laros et al. 1987, Hurley et al. 1994).  A differential energy
distribution of events follows a Gutenberg-Richter power law (-- 1.66 exponent;
Gutenberg \& Richter 1956) with a maximum energy E$_{\rm max}~\approx 
5~\times~10^{41}$ ergs (Cheng et al. 1995).  Each of these statistical
properties are consistent with characteristics of earthquakes, which suggests
the SGR bursts may be triggered by neutron star crustquakes (Thompson \& Duncan
1995).

Here, we report the discovery of SGR~1627--41, the first new SGR to be detected
since 1979.  We provide information on general burst characteristics and the
persistent X-ray emission and draw comparisons to other SGRs.  Like the other
SGRs, this SGR is associated with a persistent X-ray source and the young SNR,
G337.0--0.1.

\section{BATSE OBSERVATIONS}

During a period of intense burst activity from SGR~1900+14, the Burst and
Transient Source Experiment (BATSE; Fishman et al. 1989) trigger criteria were
optimized to detect SGR burst events.  On 1998 June 15, three consecutive BATSE
triggered bursts short in duration and having soft spectra, originated from a
region of the sky which was inconsistent with the three known SGR locations. 
Two days later, BATSE detected another 17 events from the same region,
confirming the existence of a new SGR, SGR~1627--41 (Kouveliotou et al. 1998b;
source name based upon initial BATSE location).  Over the course of the next
month and a half, a total of 99 bursts from this source were detected with
BATSE, 39 of which triggered the instrument.  Figure 1 shows the observed burst
rate as seen with BATSE.

The bursts from SGR~1627--41 last between 25 msec and 1.8 sec with most burst
durations clustering near 100 msec.  In Figure 2, we show some representative
burst profiles.  This small sample shows the diverse temporal variability
observed in bursts from this source.   The longest event (Figure 2d) is similar
to two bursts seen from SGR~1900+14 with respect to both spectrum and temporal
structure.  These bursts are much longer than typical SGR events and have very
smooth temporal profiles with abrupt $\gamma$-ray  emission start and end
points.

Due to the rapid succession of bursts on June 17 and 18, only limited fine
spectral data for a given trigger were read out from the spacecraft before the
next trigger, making detailed spectral reconstruction impossible for most
bursts.  Of the 39 triggered bursts, fine spectral resolution data were
available for only 8 events, including two very bright bursts (Figures 2c and
2d).  In fact, these two bright bursts have the two highest peak count rates of
any extra-Solar event ever observed with BATSE.  We fit power law, blackbody
and OTTB spectral models to those bursts, and find the OTTB model best
represents the time-integrated burst spectra of the six dim events and one of the two bright
bursts (Figure 2d).  The measured $kT$ values of the six dim events range
between 25 -- 35 keV; having a weighted mean of 27 keV.  The spectral
form of these six events agrees well with previous modeling of burst spectra
from the other three SGRs (see e.g. Fenimore et al. 1994).  

Due to dead-time problems at the peak of the two bright bursts, we fit spectra
taken at the tail of each event.  For the longest event (Figure 2d), we find
the tail spectrum is best fit by an OTTB model with a $kT = 27.0 \pm 0.4$ keV. 
Two spectra taken from the tail of the brightest event (Figure 2c) are
significantly harder than any other SGR~1627--41 burst spectrum measured with
BATSE.  Furthermore, they are not consistent with one another, which shows
spectral evolution exists for this event.  The OTTB and power law spectral
models cannot fit the first spectrum separately, but a combination of the two
yields an acceptable fit.  For this fit, the power law (photon) index $\alpha =
- 2.07 \pm 0.13$ and the OTTB $kT = 32 \pm 1$ keV.  The following spectrum
taken is at a lower flux level and is much harder.  We find this spectrum can
be fit by a simple power law with an index $\alpha = - 1.86 \pm 0.07$. 
Evidence for similar hard burst emission from SGR~1806--20 was found by
Strohmayer \& Ibrahim (1997).  A detailed discussion of this topic is beyond
the scope of this $Letter$, but it will be presented elsewhere (Woods et al.
1999a).

In order to estimate the peak fluxes of a larger sample of bursts, we applied
the OTTB model to bursts with coarse spectral resolution (4 channels) and fair
temporal resolution (64 msec).  We assumed a fixed $kT$ corresponding to the
measured weighted mean value for the five dim events (27 keV) and allowed only
the normalization (energy flux) to vary.  One drawback to this method is that
many bursts reach their peak flux for only a short time, less than 64 msec, so
these peak flux measurements will underestimate the true peak flux for some
events.  Given the limited data availability, however, this time scale provided
the largest sample of events.  Figure 3 shows the cumulative peak flux
distribution on the 64 msec time scale for 57 events.  The observed peak fluxes
range over 3 orders of magnitude between  $9~\times~10^{-8}$ and
$1.1~\times~10^{-4}$ ergs cm$^{-2}$ sec$^{-1}$.  Dead-time effects for the two
brightest events were excessive, so these peak flux measurements (1.1 and 0.51
$\times~10^{-4}$ ergs cm$^{-2}$ sec$^{-1}$) can be treated as lower limits. 
The dashed line represents a power law fit to this distribution which has an
exponent $\gamma = - 0.6 \pm 0.1$.  No turnover is seen for this distribution
out to $1.1~\times~10^{-4}$ ergs cm$^{-2}$ sec$^{-1}$.  We also constructed a
cumulative burst fluence distribution for these events, which has a slightly
flatter slope of $\gamma = - 0.5 \pm 0.1$.  The differential fluence (energy)
distribution then has an exponent equal to -- 1.5, which agrees well with the 
the Gutenberg-Richter power law index (-- 1.66).

%A number of factors may have artificially altered the observed cumulative peak
%flux distribution.  Dead-time problems for the brightest events would tend to
%bend the distribution down slightly at the higher peak flux values.  Selection
%of bursts by a certain threshold for triggering would result in an apparent
%deficiency of weak events.  Also, missing events due to Earth occultation of
%the source may alter the distribution for a small number of observed events. 
%The mild corrections introduced for these selection effects, however, can not
%account for the large discrepancy between this distribution and the
%Gutenburg-Richter power law.

Using the BATSE triggers, the burst source was coarsely located (Kouveliotou et
al. 1998b) at $\alpha$ = 16$^{\rm h}$ 27$^{\rm m}$ and $\delta$ =
-- 41$^{\circ}$ (J2000) with an error circle of radius 2$^{\circ}$.  Detection
of SGR events by both BATSE and the Ulysses spacecraft provided a narrow
location annulus 1.7$^{\prime}$ wide (Hurley et al. 1998a).  Using BATSE Earth
occultation constraints, we limited the allowable range along the annulus to
1.5$^{\circ}$ (Woods et al. 1998; Figure 4).  A more detailed account of the
localization of this SGR is reported in Hurley et al. (1999d) and Smith et al.
(1999).  In view of the association of SGRs with young SNRs, we searched the
Whiteoak \& Green (1996) catalogue of SNRs near the refined error box.  A
single SNR, G337.0--0.1 (Sarma et al. 1997), was found (Woods et al. 1998)
within the 1.5$^{\circ} \times 1.7^{\prime}$ error box.  With hopes of
detecting an X-ray counterpart for this SGR, a ToO observation of this SNR was
initiated using the BeppoSAX (Boella et al. 1997a) Narrow Field Instruments
(NFIs).

\section{BeppoSAX OBSERVATIONS}

Two observations of SNR G337.0--0.1 were performed on 1998 August 7 and again
on September 16.  These observations revealed a previously undetected X-ray
source (SAX~J1635.8--4736) at $\alpha$ = 16$^{\rm h}$ 35$^{\rm m}$ 49.8$^{\rm
s}$ and $\delta$ = --47$^{\circ}$ 35$^{\prime}$ 44$^{\prime \prime}$ (J2000)
with an error circle of radius 1$^{\prime}$ (95\% confidence; Figure 4),
consistent with the SNR location.  A known source, 4U1630--472, is also seen
near the edge of the field of view for each observation.  A light curve of the
new source for each observation does not show any burst activity which is
consistent with BATSE observations for those time periods (see Figure 1). 
Using the Low Energy Concentrator Spectrometer (LECS; Parmar et al. 1997) and
two Medium Energy Concentrator Spectrometers (MECS; Boella et al. 1997b), we
fit the spectrum of SAX~J1635.8--4736 from 0.1 -- 10 keV.  The spectrum is well
represented by a power law with interstellar absorption.  Under the assumption
that the spectral form (i.e. the power law index and Hydrogen column density)
remains constant between observations, we fit the observations simultaneously
allowing only the normalization to vary between the two.  We get an acceptable
fit with a reduced $\chi^2$ value of 0.92 for 160 degrees of freedom and find a
power law (photon) index $\alpha$ = --~2.5~$\pm$~0.2 and a column density
N$_{\rm H}$ =  (7.7~$\pm~0.8)~\times~10^{22}$~cm$^{-2}$.  The unabsorbed flux
(2 -- 10 keV) declines between the observations (40.3 days) from
$(6.7~\pm~0.3)~\times~10^{-12}$ ergs~cm$^{-2}$~sec$^{-1}$ to
$(5.2~\pm~0.4)~\times~10^{-12}$ ergs~cm$^{-2}$~sec$^{-1}$. Assuming this source
is located within the SNR, the distance is 11 kpc (Sarma et al. 1997). The
source luminosity is then  9.7~$\times$~10$^{34}$~ergs~sec$^{-1}$ and 
7.6~$\times$~10$^{34}$~ergs~sec$^{-1}$ for the two observations.  

%Using the fact that 4U1630--472 is detected during our observations of SNR
%G337.0--0.1, we searched an archival SAX NFI observation of 4U1630--472 from
%March 1997 in order to see if SAX~J1635.8--4736 is visible.  We find this
%source %is not visible and the 2$\sigma$ upper limit on the observed flux is
%$7~\times~10^{-12}$ ergs~cm$^{-2}$~sec$^{-1}$, assuming the measured spectral
%index and column density.

Using standard SAX analysis techniques, source counts were extracted from the
combined MECS units for SAX~J1635.8--4736 and binned at 0.5 sec time
resolution.  We then performed a Fast Fourier transform (FFT) of the 1998
August light curve searching frequencies from 0 -- 1 Hz and found the largest
value in the power density spectrum was at 0.156 Hz.  Although not very
significant by itself, the corresponding period falls within the tight range of
observed periods (5 -- 8 sec) for the other SGRs.  Using the barycenter
corrected time tags, we ran an epoch fold search about the period corresponding
to the highest power, which revealed a marginally significant peak
(6~$\times~10^{-3}$ chance probability taking into account the number of
trials; 1500 between 6.38 and 6.44 sec) at 6.41318(3) sec (JD = 2451032.5). 
The nearly sinusoidal pulse profile has an r.m.s. pulse fraction of
10.0~$\pm$~2.6~\%.  We performed the same analysis on the 1998 September
observation, but did not find any significant peak in the power density
spectrum near 0.156 Hz or anywhere else between 0 -- 1 Hz.  However, given the
weak signal found in August and the fact that 45\% fewer source counts were
recorded in the MECS units during the September observation, we would not
expect to find this pulsed signal.

%\section{RXTE OBSERVATIONS}

%Shortly after the discovery of SGR~1627--41 and before the SAX ToO
%observations were performed, a ToO of the region near G337.0--0.1 was
%initiated using the {\it Rossi X-ray Timing Explorer} Proportional Counter
%Array (PCA; Jahoda et al. 1997).  Data were collected from June 26 with 9.2
%ksec of on-source time, during which a single burst was detected.  A power
%spectral analysis of these data does not reveal coherent pulsations in the
%range 0 -- 1 Hz.  However, a quasi-periodic oscillation (QPO) is detected with
%a centroid frequency of 0.15 Hz (Dieters et al. 1998).  As mentioned earlier,
%4U1630--472, a black hole transient, is clearly visible during the SAX
%observations in August and September.  In fact, the combined MECS count rate
%is larger that that of SAX J1635.8--4736 in August by a factor 20.  If we
%assume a linear flux decay for SAX~J1635.8--4736 between August and September,
%then extrapolate backwards to June 26, we find that the expected count rate in
%the PCA for SAX~J1635.8--4736 would be only $\sim$ 2.6 c/s.  4U1630--472 was
%within the field-of-view of the PCA in June with a much higher count rate and
%was found to display QPO from  0.1 and 1.0 Hz during 1998 May 20 to 1998 June
%8 (Tomsick 1999).  Therefore, we conclude that the QPO at 0.15 Hz, although
%close in frequency to the weak periodic signal found in the SAX data, most
%likely originates from the black hole transient.

\section{DISCUSSION}

We propose that SAX~J1635.8--4736 is the X-ray counterpart to SGR~1627--41. 
There are a number of observations discussed here which support this claim. 
First, the position of SAX~J1635.8--4736 is mutually consistent with the narrow
error box for SGR~1627--41 (Hurley et al. 1999d, Smith et al. 1999) and the SNR
G337.0--0.1.  Second, its spectrum is very similar to those found for the other
SGR X-ray counterparts (Hurley et al. 1999a).  Also, this X-ray source is
variable near a burst active period for SGR~1627--41 which has also been found
for the X-ray counterpart of SGR~1900+14 (Hurley et al. 1999a, Murakami et al.
1999, Woods et al. 1999b).  Finally, the marginal detection of pulsations at
6.4 sec, if confirmed, would agree well with the known spin periods of the
three other SGRs which fall within a tight range of 5 -- 8 sec (Kouveliotou et
al. 1998a, Hurley et al. 1999a, Mazets et al. 1979).

For the distance of 11 kpc, G337.0--0.1 has a small diameter of $\sim$ 5 pc
(Sarma et al. 1997).  There are only a few SNRs this small (see Case \&
Bhattacharya 1998), and a large fraction of them are very young (e.g. Tycho,
Kepler, Cas A).  This suggests that G337.0--0.1 is also very young.  The
association of SGR~1627--41 with G337.0--0.1 strengthens the connection between
SGRs and young SNRs.

Given the distance to the SNR G337.0--0.1 and assuming isotropic emission, the
burst peak luminosities ($>$ 25 keV) vary from 10$^{39}$ to 10$^{42}$ ergs
sec$^{-1}$ (Figure 3).  Paczynski (1992) suggested that SGR~0526-66 may have a
critical luminosity $\sim$ $2~\times~10^{42}$ ergs sec$^{-1}$.  He calculated
the relation between the dipole magnetic field (B$_{\rm dipole}$) of a neutron
star and the critical luminosity (L$_{\rm crit}$) the magnetosphere will allow
to escape in the limit where L$_{\rm crit}~\gg$~L$_{\rm Edd}$ (L$_{\rm Edd}$
is the standard Eddington luminosity for a 1.4 $M_{\odot}$ neutron star).  This
relation is given by 

\begin{displaymath}
   \frac{\rm L_{crit}}{\rm L_{Edd}} \approx 2 \left( \frac{\rm B_{dipole}} 
   {10^{12} {\rm Gauss}} \right) ^{4/3} \left( \frac{g}{2 \times 10^{14} 
   {\rm cm~sec}^{-1}} \right) ^{-1/3}
\end{displaymath}

\noindent where $g$ is the surface acceleration due to gravity. For
SGR~0526--66, Paczynski found a dipole magnetic field of $\sim$
$6~\times~10^{14}$ Gauss, which agrees with independent estimates made by
Duncan \& Thompson (1992).  For SGR~1627--41, we do not detect a turnover in
the cumulative peak luminosity distribution, but we can place a lower limit on
this value of $1.6~\times~10^{42}$ ergs sec$^{-1}$ based upon the highest
observed peak luminosity.  This, in turn, places a lower limit on the magnetic
field of the presumed magnetar associated with SGR~1627--41 of B$_{\rm
dipole}~\gtrsim 5~\times~10^{14}$ Gauss.  This estimate may be confirmed with
the definitive measurement of a pulse period and its derivative for
SGR~1627--41.

\acknowledgments{\noindent {\it Acknowledgements} -- We thank other BATSE team
members for useful discussions of dead-time effects of the instrument.  This
research has made use of SAXDAS linearized and cleaned event files produced at
the BeppoSAX Science Data Center.  We would also like to thank Lorella Angelini
for her assistance with analyzing the SAX data.  The MOST is operated by the
University of Sydney with support from the Australian Research Council and the
Science Foundation for Physics within the University of Sydney. P.M.W.
acknowledges support from NASA through grant NAG 5-3674.  J.v.P. acknowledges
support from NASA through grants NAG 5-3764, NAG 5-7060 and NAG 5-7808.}

\newpage

\figcaption{Burst rate of SGR~1627--41 (triggered and untriggered) as observed
	with BATSE.  In total, 99 bursts were detected within 60 days.  Arrows
	indicate times of BeppoSAX NFI observations of G337.0--0.1.} 

\figcaption{A sample of six bursts of SGR~1627--41 observed with the BATSE
	Large Area Detectors (LADs).  The upper and lower panels are
	Time-Tagged Event (TTE) data ($>$ 25 keV) accumulated with 4 msec (a),
	2 msec (b), 2 msec (e) and 1 msec (f) time resolution.  The middle
	panels are DISCriminator SCience (DISCSC) data ($>$ 25 keV) accumulated
	with 64 msec time resolution.  Trigger times for these six events are
	50981.87322 (a), 50982.03545 (b), 50982.07120 (c), 50982.16969 (d),
	50993.30911 (e) and 51006.91017 (f) in MJD (UT).  BATSE trigger
	numbers are given in the upper left corner of each panel.} 

\figcaption{Cumulative peak flux (0.064 sec) distribution for 57 events. 
	Dashed line is a power law fit to these data having an exponent equal
	to -- 0.6.  Bottom horizontal axis labels the peak flux and the top
	horizontal axis gives the peak luminosity assuming isotropic emission
	and an 11 kpc distance to the source.} 

\figcaption{Localization of SGR~1627--41 and SAX~J1635.8--4736.  Panel (a)
	shows the Interplanetary Network (IPN) arc ({\it solid lines}) and
	Earth occultation constraints ({\it shaded region}).  Panel (b) is a
	magnification of region near G337.0--0.1 (radio contours; Whiteoak \&
	Green 1996).  Solid straight lines represent BATSE-Ulysses IPN arc. 
	Dotted lines are the Konus-Ulysses IPN arc (Hurley et al. 1998b). 
	Solid, bold circle represents error region for SAX J1635.8-4736.}

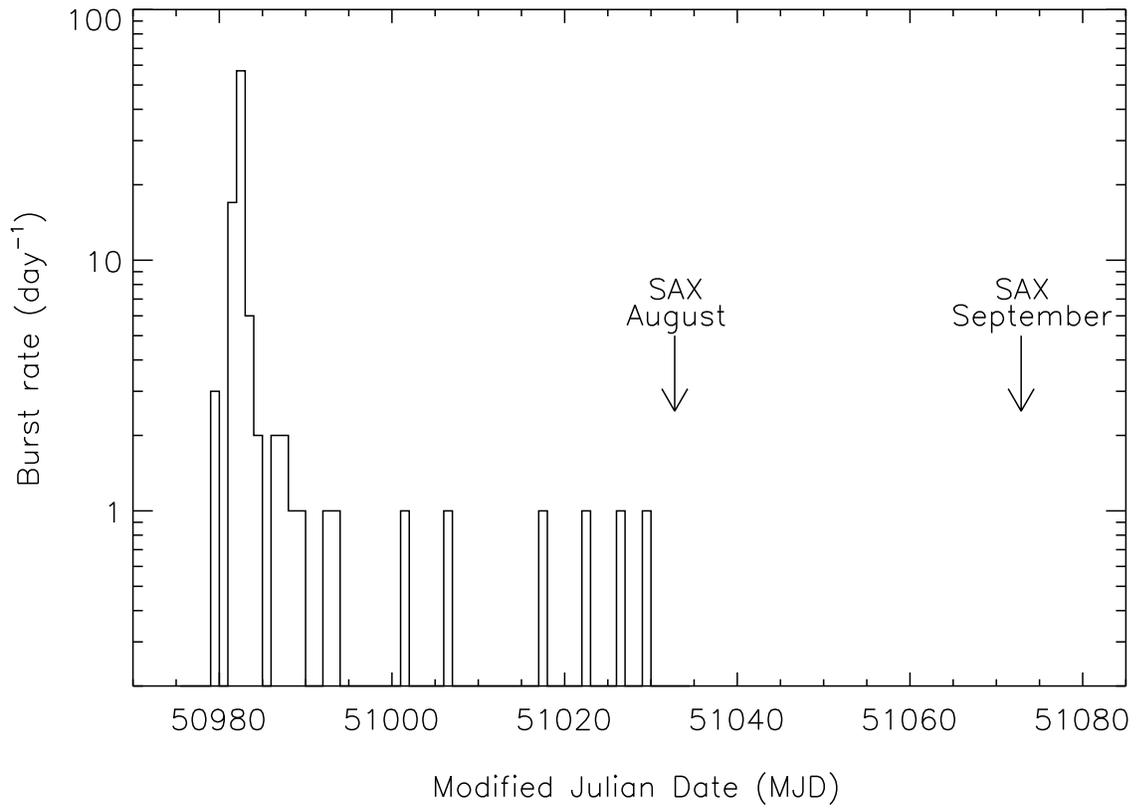
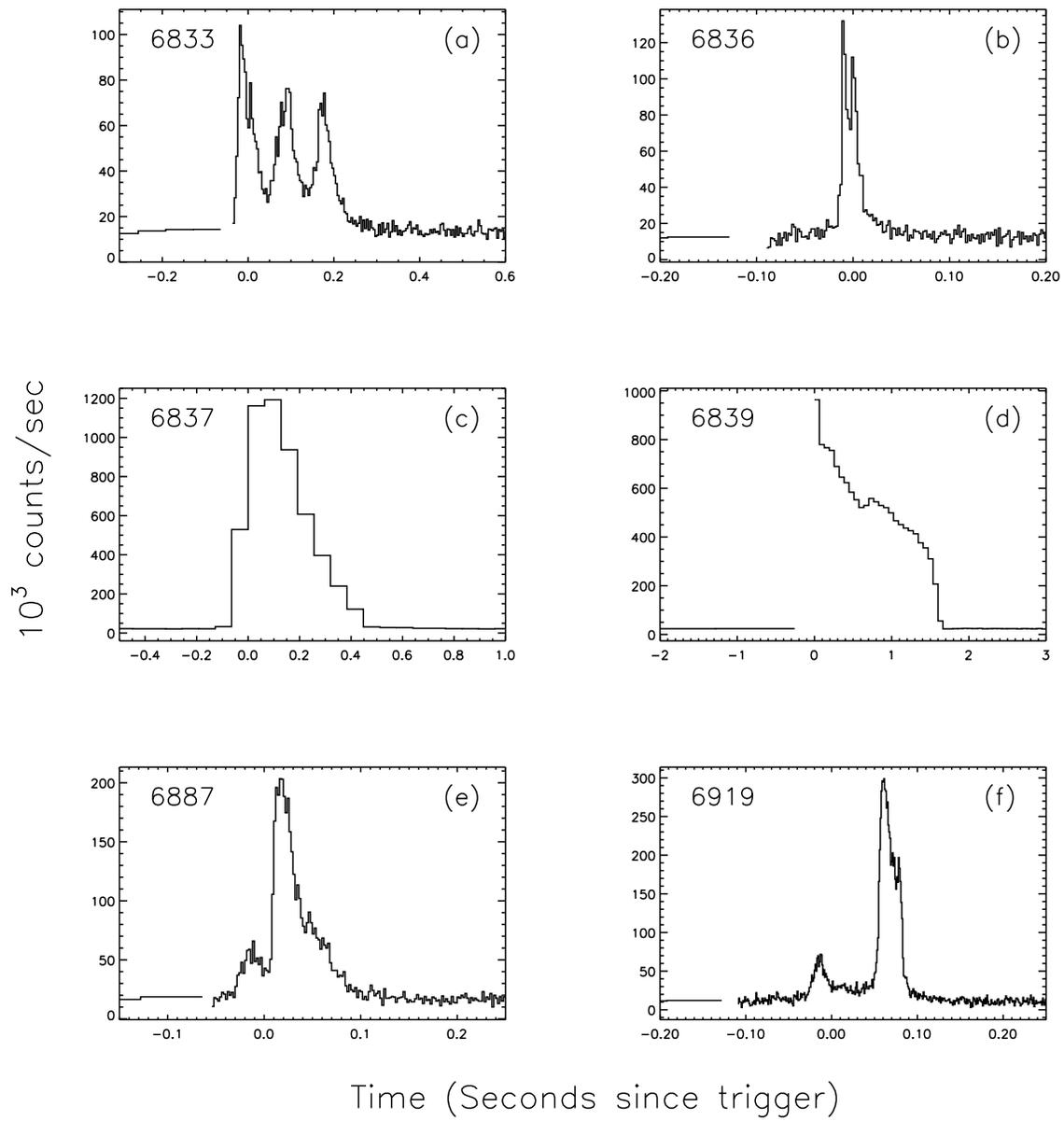
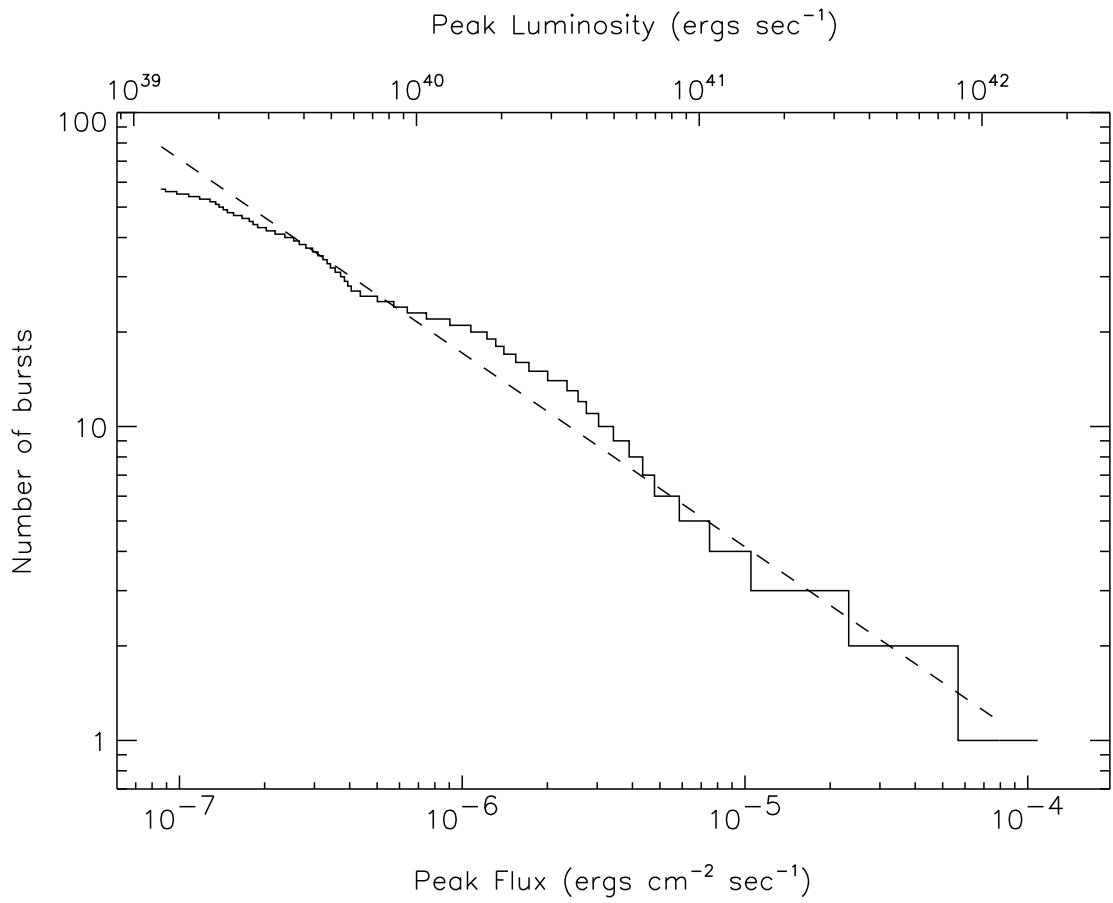
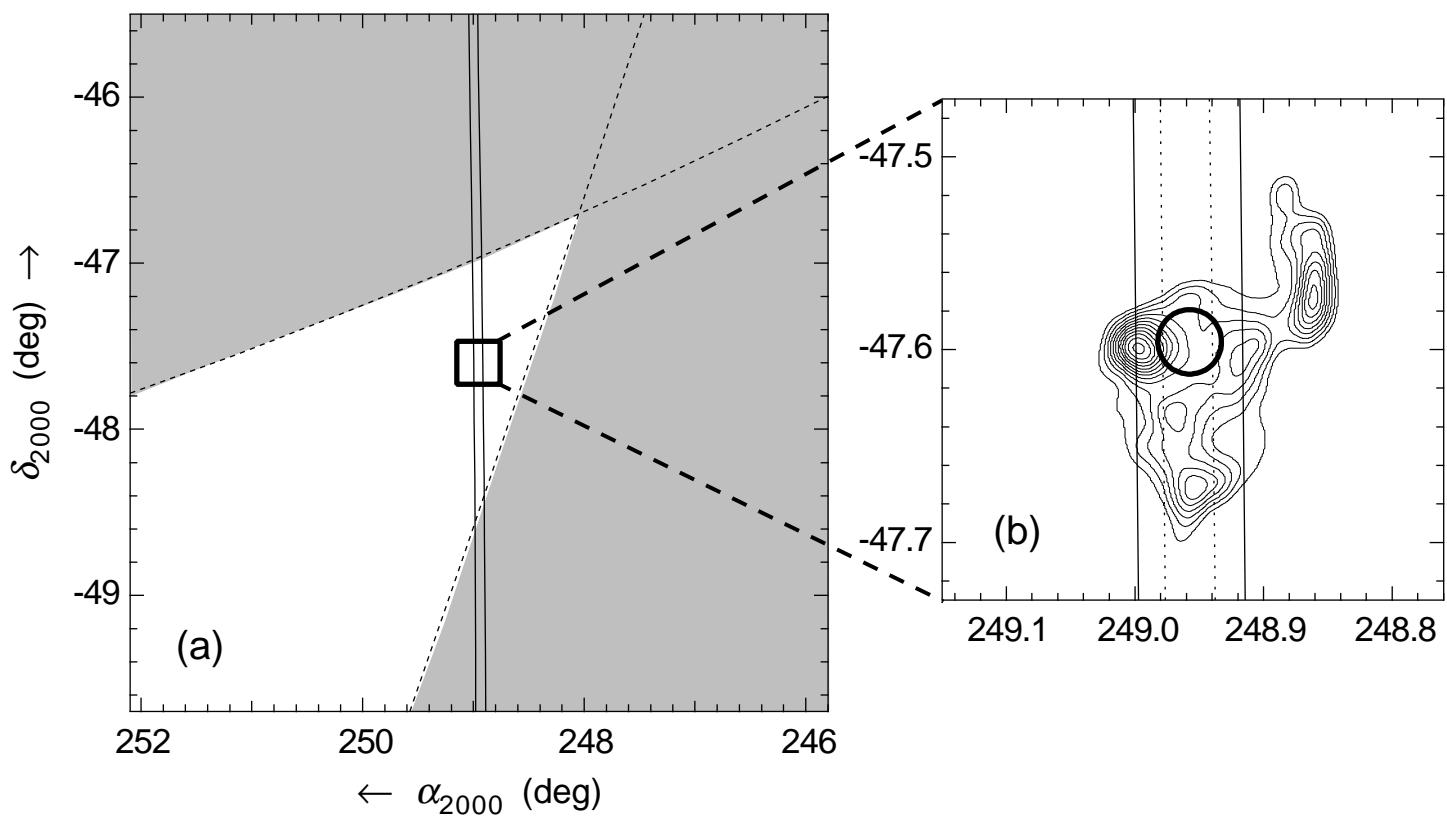
\end{document}